\documentclass{ws-ijmpe}
\usepackage[super,compress]{cite}

\usepackage{subfigure}

\begin{document}
%

\catchline{}{}{}{}{}

\title{Universal descriptions of chemical freeze-out based on pressure and specific heat respectively
}

\author{\footnotesize Subhasis Samanta $^1$ $^2$
}

\address{$^1$School of Physical Sciences, National Institute of Science 
Education and Research, HBNI, Jatni - 752050, India
\\
subhasis.samant@gmail.com}

\address{$^2$Center for Astroparticle Physics \&
Space Science, Bose Institute, Block-EN, Sector-V, Salt Lake, Kolkata-700091, India 
 \\ \& \\ 
Department of Physics, Bose Institute, \\
93/1, A. P. C Road, Kolkata - 700009, India}

\maketitle

\begin{history}
\received{Day Month Year}
\revised{Day Month Year}
\end{history}

\begin{abstract}
The lattice QCD data of pressure and the energy density have been used to extract the 
hadronic radius parameter of the excluded volume hadron resonance gas (EVHRG) model.
The equation of state can be described well with the extracted radius parameter $R_h= 0.15$ fm.
Specific heat is also calculated in the EVHRG model. 
Further, two new universal descriptions of chemical freeze-out parameters
have been introduced based on pressure and specific heat respectively.
It is shown that the chemical freeze-out parameters obtained 
at various $\sqrt{s_{NN}}$ in ideal HRG model approximately 
correspond to $P/T^4= 0.88$ and $C_V/T^3= 47$ respectively.
These two quantities are important to describe
 the thermodynamic properties of the hadronic matter created in heavy ion collision experiment.
The sensitivity of universal chemical freeze-out lines on repulsive interaction
is also studied.
It has been observed that the behaviors of chemical freeze-out lines for
$P/T^4$ and $C_V/T^3$ in EVHRG model remain similar to ideal HRG model
for the best fit value of hadronic radii.
\end{abstract}


\ccode{PACS numbers:25.75.-q; 24.10.Pa}

\section{\label{sec:Intro} Introduction}
In the last few years, a substantial amount of 
experimental and theoretical efforts have been devoted
worldwide to investigate the strongly interacting matter 
under extreme conditions of temperature and/or baryon chemical potential.
While at low baryon chemical potential and high temperature
lattice quantum chromo dynamics (LQCD) data seem to indicate a smooth
crossover from hadronic to quark-gluon plasma (QGP) phase ~\cite{Aoki:2006we}, at high 
baryon chemical potential and low temperature the system is expected to have a
first-order phase transition ~\cite{Asakawa:1989bq}. 
This first order phase transition line at high baryon chemical potential and low temperature 
should end at a critical end point (CEP), a second-order phase transition point as
one moves towards the high temperature and low baryon chemical potential region, in the QCD phase diagram.
Heavy ion collisions provide a unique tool to create and study
strongly interacting matter under extreme conditions of temperature and/or baryon chemical potential.
One of the primary goals of heavy-ion collision experiments is to map QCD phase diagram
in terms of temperature and baryon chemical potential.
At present, the properties of QCD matter at high temperature
and almost zero baryon chemical potential are being investigated using ultra relativistic 
heavy ion collisions at the Large Hadron Collider (LHC), CERN
and Relativistic Heavy Ion Collider (RHIC), Brookhaven National Laboratory (BNL).
The Beam Energy Scan (BES) program of RHIC is currently
investigating the location of the CEP.
The HADES experiment
at GSI, Darmstadt is investigating medium of very large baryon chemical potential.
The region of large baryon chemical potential will also be explored by the NA61-SHINE experiment at CERN-SPS.
In future, the Compressed Baryonic Matter (CBM) experiment at the Facility for Antiproton and
Ion Research (FAIR) at GSI and the Nuclotron-based Ion Collider fAcility (NICA) 
at JINR, Dubna will also study nuclear matter at large baryon chemical potential. 

LQCD provides the most direct approach to study the QCD matter at high temperature.
However, at finite
chemical potential, LQCD faces the well-known sign problem.
On the other hand,
the hadron resonance gas (HRG) model ~\cite{Hagedorn:1980kb, Rischke:1991ke,%
Cleymans:1992jz, BraunMunzinger:1994xr, Cleymans:1996cd, Yen:1997rv, BraunMunzinger:1999qy,%
Cleymans:1999st, BraunMunzinger:2001ip, BraunMunzinger:2003zd, Karsch:2003zq, Tawfik:2004sw,%
Becattini:2005xt, Andronic:2005yp, Andronic:2008gu,Begun:2012rf, Andronic:2012ut,%
Tiwari:2011km, Fu:2013gga, Tawfik:2013eua, Garg:2013ata, Bhattacharyya:2013oya,%
Bhattacharyya:2015zka,Kadam:2015xsa, Kadam:2015fza, Albright:2014gva,%
Albright:2015uua,Bhattacharyya:2015pra, Begun:2016cva,Adak:2016jtk, Xu:2016skm,Fu:2016baf}, 
which is used in this present work,
provide a simpler
alternative to study strongly interacting matter 
at finite temperature and chemical potential.
The HRG model is quite successful in describing the bulk properties of 
hadronic matter in thermal and chemical
equilibrium ~\cite{Karsch:2003zq, Tawfik:2004sw, Andronic:2012ut}.
This model is also successful in describing the ratios of hadron yields, 
at chemical freeze-out, created in central 
heavy ion collisions from SIS up to RHIC energies~~\cite{BraunMunzinger:1994xr, Cleymans:1996cd, 
BraunMunzinger:1999qy, Cleymans:1999st, BraunMunzinger:2001ip, Becattini:2005xt, 
Andronic:2005yp, Andronic:2008gu}. In a heavy ion collision experiment, the chemical freeze-out is defined as the
stage in the evolution of the thermal system when inelastic
collisions among the hadrons cease and the hadronic ratios become fixed.
At various center-of-mass energies $\sqrt{s_{NN}}$, 
hadronic yields or ratios are generally analyzed phenomenologically using HRG model
~\cite{Cleymans:2005xv,Xu:2001zj,Becattini:2005xt,Andronic:2005yp,Andronic:2009jd,
Karsch:2010ck,Chatterjee:2015fua} to determine the chemical freeze-out parameters.
Relations of chemical freeze-out temperature
and baryon chemical potential with $\sqrt{s_{NN}}$ establish
the chemical freeze-out line ~\cite{Cleymans:2005xv,Andronic:2005yp}
in the QCD phase diagram.
The main result of these investigations is that the chemical freeze-out temperature 
increases sharply from SIS up to SPS energies and reaches, for
higher collision energies, at constant
values near $T = 160-165$ MeV while baryon chemical potential decreases sharply
as a function of $\sqrt{s_{NN}}$. Chemical freeze-out is the earliest stage in the evolution of the hadronic phase
which can be determined phenomenologically from the experiment data. 
Therefore, the chemical freeze-out line is very much important
in the QCD phase diagram. 
To know the thermodynamic properties of the system at chemical freeze-out
we have to study equation of state of the matter.
There are several universal chemical freeze-out descriptions in the 
existing literature which can approximately describe the chemical freeze-out
line in the QCD phase diagram. Those universal properties will be discussed in detail in this paper.
Those universal chemical freeze-out descriptions are independent of $\sqrt{s_{NN}}$
and they are related to the equation of state of the thermal system.
Those universal properties will be useful to study properties of thermal system created 
in heavy-ion collisions at any $\sqrt{s_{NN}}$.
Finding out the universal
conditions of chemical freeze-out parameters have been the subject of various studies.

The aim of the present work is twofold. 
First, I would like to study some basic bulk thermodynamic
quantities, where all the hadrons in HRG model contribute,
like pressure, energy density, entropy density and specific heat
of the matter in the HRG and an EVHRG model. 
Second, I would like to find out universal
conditions of chemical freeze-out descriptions from those
thermodynamic quantities.

The paper is organized as follows.
First, the ideal HRG model and an EVHRG model are briefly discussed in Sec. \ref{sec:HRG}.
The Sec. \ref{sec:results} shows results of this paper.
In Sec. \ref{sec:EOS_mu0} the effect of repulsive interaction at zero chemical potential
has been studied using the LQCD data of equation of state. Further, in Sec. \ref{sec:cv_mu0}
result of specific heat at $\mu = 0$ has been shown.
Thereafter, two
new universal chemical freeze-out descriptions based on pressure and specific heat respectively
have been proposed in the Sec. \ref{sec:p_cv_cfo}. Furthermore, the behavior of universal 
chemical freeze-out lines 
in presence of repulsive interaction has been studied in Sec. \ref{sec:sensitivity}.
Finally in the Sec. \ref{sec:conclusions}, I conclude
the findings of the paper.

\section{\label{sec:HRG} Hadron Resonance Gas model}
In the HRG model, the thermal system consists of 
all the hadrons and resonances. There are varieties of HRG models in the existing literature.
Different versions of this model and some of the recent works using these models
may be found in Refs. ~\cite{Hagedorn:1980kb, Rischke:1991ke,%
Cleymans:1992jz, BraunMunzinger:1994xr, Cleymans:1996cd, Yen:1997rv, BraunMunzinger:1999qy,%
Cleymans:1999st, BraunMunzinger:2001ip, BraunMunzinger:2003zd, Karsch:2003zq, Tawfik:2004sw,%
Becattini:2005xt, Andronic:2005yp, Andronic:2008gu,Begun:2012rf, Andronic:2012ut,%
Tiwari:2011km, Fu:2013gga, Tawfik:2013eua, Garg:2013ata, Bhattacharyya:2013oya,%
Bhattacharyya:2015zka,Kadam:2015xsa, Kadam:2015fza, Albright:2014gva,%
Albright:2015uua,Bhattacharyya:2015pra,Vovchenko:2015cbk, Begun:2016cva,Adak:2016jtk, Xu:2016skm,Fu:2016baf,
Vovchenko:2016ebv,Satarov:2016peb, Vovchenko:2016eby, Oliinychenko:2016dtb,Tawfik:2016jzk}.
The logarithm of the partition function of a hadron resonance gas in the grand canonical ensemble can be written as 
\begin {equation}
 \ln Z^{id}=\sum_i \ln Z_i^{id},
\end{equation}
where the sum is over all the hadrons and resonances, $id$ refers to ideal {\it i.e.}, non-interacting HRG model.
For hadron species $i$,
\begin{equation}
 \ln Z_i^{id}=\pm \frac{Vg_i}{2\pi^2}\int_0^\infty p^2\,dp \ln[1\pm\exp(-(E_i-\mu_i)/T)],
\end{equation}
where $V$ is the volume of the thermal system, $T$ is the temperature, $g_i$ is the degeneracy factor,
$m_i$ is the mass,
$E_i(p)=\sqrt{{p}^2+m^2_i}$ is the single particle energy and $\mu_i=B_i\mu_B+S_i\mu_S+Q_i\mu_Q$ is the
chemical potential. In the last expression, $B_i,S_i,Q_i$ are respectively
the baryon number, strangeness and electric charge of the hadron, $\mu^,s$ are corresponding chemical 
potentials. 
The upper and lower
sign corresponds to baryons and mesons respectively.
In this present work, all the hadrons and resonances listed in the particle 
data book up to a mass of 3 GeV ~\cite{Agashe:2014kda} have been incorporated.
The width of the resonances are taken as zero. In this approximation a
resonance behaves identically to that of a stable hadron \cite{Dashen:1969ep}.
Hence resonance decay is also not there in this model.
It is assumed that the hadronic matter is
in thermal and chemical equilibrium. 
The partition function is the basic quantity from which one can calculate various thermodynamic 
quantities of the thermal system.
The pressure $P^{id}$, energy density
$\varepsilon^{id}$, entropy density $s^{id}$ and the number density $n^{id}$ of the thermal system
can be calculated using the standard definitions,

\begin{align}\label{eq:p}
 \begin{split}
P^{id} &=\sum_i P_i^{id}=\sum_i T\frac{\partial \ln Z_i^{id}}{\partial V}\\
 & =\sum_i \pm\frac{g_iT}{2\pi^2}\int_0^\infty p^2\,dp \ln[1\pm\exp(-(E_i-\mu_i)/T)],
 \end{split}
\end{align}

\begin{align}\label{eq:e}
\begin{split}
\varepsilon^{id}=\sum_i \varepsilon_i^{id}&=- \sum_i  \frac{1}{V} \left(\frac{\partial \ln Z_i^{id}}{\partial\frac{1}{T}}\right)_{\frac{\mu}{T}}\\
&=\sum_i \frac{g_i}{2\pi^2}\int_0^\infty\frac{p^2\,dp}{\exp[(E_i-\mu_i)/T]\pm1}E_i,
\end{split}
\end{align}

\begin{align}\label{eq:s}
 \begin{split}
  s^{id}&=\sum_i \frac{1}{V}\left(\frac{\partial\left({T \ln Z_i^{id}}\right)}{\partial T}\right)_{V,\mu}\\
& =\sum_i \pm\frac{g_i}{2\pi^2}\int_0^\infty p^2\,dp \left[ \ln\left(1\pm\exp(-\frac{(E_i-\mu_i)}{T})\right)\right.
 \left.\pm\frac{(E_i-\mu_i)}{T(\exp((E_i-\mu_i)/T)\pm1)}\right].
 \end{split}
\end{align}

\begin{align}\label{eq:n}
\begin{split}
 n^{id} &=\sum_i  \frac{T}{V} \left(\frac{\partial \ln Z_i^{id}}{\partial\mu_i}\right)_{V,T}\\
& =\sum_i  \frac{g_i}{2\pi^2}\int_0^\infty\frac{p^2\,dp}{\exp[(E_i-\mu_i)/T]\pm1}.
\end{split}
\end{align}

In case of heavy-ion collision experiments, the parameters $T$ and $\mu's$ of
HRG model corresponds to those at
chemical freeze-out which are believed to depend on initial conditions of the collision
due to global charge conservation. That means conserved initial charges of the system
will not change even after the collision.
The chemical potentials at chemical freeze-out $\mu_B, \mu_S$ and $\mu_Q$
are not independent but related (on average) to each other as well as to the $T$
via the relations ~\cite{Alba:2014eba}
\begin{equation}
\label{eq:ns}
\sum_i n_i (T, \mu_B, \mu_S, \mu_Q) S_i=0,
\end{equation}
and
\begin{equation}
\label{eq:nbq}
 \sum_i n_i (T, \mu_B, \mu_S, \mu_Q) Q_i= r \sum_i n_i (T, \mu_B, \mu_S, \mu_Q) B_i,
\end{equation}
where $n_i$ is the number density of $i$ th hadron at chemical freeze-out,
$r$ is the ratio of net-charge to net-baryon number of the colliding nuclei.
For central Au + Au or Pb +Pb collisions $r = N_p /(N_p + N_n)=0.4$, where $N_p$ and $N_n$ are 
respectively proton numbers and neutron numbers of the colliding nuclei.
The right-hand side of the Eq. \ref{eq:ns} is zero since initially there is no
net-strangeness in the colliding nuclei. Similarly Eq. \ref{eq:nbq} is due to the conservation
of electric charge and baryon number.

\subsection{\label{sec:EVHRG}Excluded Volume Hadron Resonance Gas model}
In the ideal HRG model hadrons and resonances are point-like.
Although attractive interactions between hadrons are incorporated through the
presence of resonances, repulsive interactions are ignored completely in this framework.
However, the repulsive interactions are also needed, especially at very high temperature
and/ or large baryonic chemical potential,
to catch the basic qualitative features of
strong interactions where the ideal gas assumption becomes inadequate.
In the EVHRG model 
~\cite{Hagedorn:1980kb, Rischke:1991ke, Cleymans:1992jz, Yen:1997rv, 
Begun:2012rf, Andronic:2012ut, Fu:2013gga, Tawfik:2013eua, Bhattacharyya:2013oya, 
 Albright:2014gva,Vovchenko:2014pka, Albright:2015uua,Kadam:2015xsa, Kadam:2015fza,
 Vovchenko:2015cbk,
 Kapusta:2016kpq, Andronic:2016nof, Vovchenko:2016ebv,Satarov:2016peb,Vovchenko:2016eby,Tawfik:2016jzk}, 
hadronic phase is modeled by a gas
of interacting hadrons, where the geometrical sizes of the
hadrons are explicitly incorporated as the excluded volume
correction to approximate the short-range van der Waals type repulsive interaction.
Excluded volume corrections were first introduced in Ref.~\cite{Hagedorn:1980kb}
but it was thermodynamically inconsistent. A thermodynamically consistent excluded volume correction was first
proposed in Ref.~\cite{Rischke:1991ke}.

In a thermodynamically consistent EVHRG model pressure can be written as \cite{Rischke:1991ke,Yen:1997rv,Andronic:2012ut}
\begin{equation}\label{eq:p_new}
 P(T,\mu_1,\mu_2,..)=\sum_i P_i^{id}(T,\hat{\mu}_1,\hat{\mu}_2,..),
\end{equation}
where for the $i$-th hadron the effective chemical potential is
\begin{equation}\label{eq:mu}
 \hat{\mu_i}=\mu_i-V_{ev,i}P(T,\mu_1,\mu_2,..)
\end{equation}
where
$V_{ev,i}=4 \, \frac{4}{3} \, \pi R_i^3$ is the volume excluded for
that hadron with hardcore radius $R_i$. 
In an iterative procedure, one can get the pressure. 
In this work we consider same radii $R_h$ for all the hadrons.
The pressure $P(T,\mu_1,\mu_2,..)$ is suppressed compared to 
the ideal gas pressure $P^{id}$ because of the smaller value of effective chemical potential. The other
thermodynamic quantities like  $\varepsilon$, $s$ and $n$
can be calculated from Eqs. \ref{eq:p_new} - \ref{eq:mu} as

\begin{equation}\label{eq:e_ev}
 \varepsilon=\varepsilon(T,\mu_1,\mu_2,..)=\frac{\sum_i \varepsilon_i^{id}(T,\hat{\mu_i})}{1+\sum_k V_{ev,k}n_k^{id}(T,\hat{\mu_k})},
\end{equation}

\begin{align}
 \begin{split}
  s=s(T,\mu_1,\mu_2,..)&=\left(\frac{\partial P}{\partial T}\right)_{\mu_1,\mu_2,..}
 =\frac{\sum_i s_i^{id}(T,\hat{\mu_i})}{1+\sum_k V_{ev,k}n_k^{id}(T,\hat{\mu_k})},
 \end{split}
\end{align}

\begin{align}\label{eq:n_ev}
 \begin{split}
   n=\sum_i n_i(T,\mu_1,\mu_2,..)
  =\frac{\sum_i n_i^{id}(T,\hat{\mu_i})}{1+\sum_k V_{ev,k}n_k^{id}(T,\hat{\mu_k})}.
 \end{split}
\end{align}
This EVHRG model is thermodynamically consistent $i.e.,$
equation of state after excluded volume corrections obey the relation 
\begin{equation}
\varepsilon + P -\sum_i \mu_i n_i = Ts. 
\end{equation}
Recently, effects of excluded-volume have been studied in 
the equation of state of pure Yang-Mills theory ~\cite{Alba:2016fku}.

\section{\label{sec:results}Results}
\subsection{\label{sec:EOS_mu0}Equations of state at $\mu =0$}

\begin{figure}[]
\centering
 \includegraphics[width=0.45\textwidth]{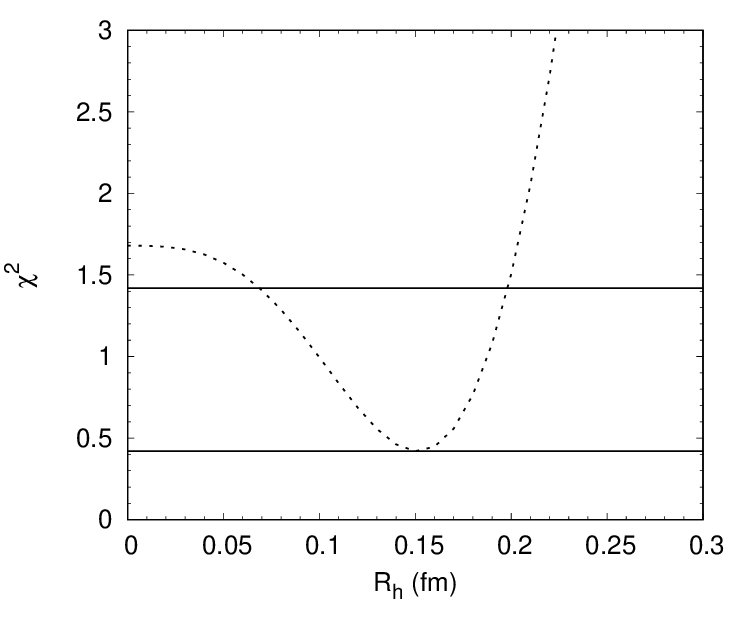}
 \vspace{0.5cm}
\caption{The variation of $\chi^2$ with the hardcore radii of hadrons ($R_h$)
at $\mu=0$.
LQCD data of ~\cite{Bazavov:2014pvz} have been used to calculate $\chi^2$.
Horizontal lines in this plot correspond to the  $\chi^2_{min}$ and the $\chi^2_{min} + 1$ values.}
\label{fig:chi2}
\end{figure}

The difference of the EVHRG model, as compared to HRG,
is governed by the radius parameter.
In this work I have tried to estimate the value of hardcore radius parameter
by fitting LQCD data of equation of state instead of tuning $R_h$ by hand.
Continuum limit LQCD data ~\cite{Bazavov:2014pvz} of two independent
thermodynamic observables of the equation of state namely
normalized pressure and energy density 
calculated at $\mu=0$ have been used to understand the effect repulsive interaction in terms of
hard core radii of hadrons in EVHRG model.
Using those LQCD data,
$\chi^2$ has been calculated at different radii of hadrons $R_h$ where
$\chi^2$ has been defined as
\begin{align}\label{eq_chi2}
 \begin{split}
   \chi^2&=\frac{1}{N} \sum_i [\frac{ \left((\frac{P}{T^4})_i^{LQCD}-(\frac{P}{T^4})_i^{EVHRG}(R_h)\right)^2}{( (\Delta\frac{P}{T^4})_i^{LQCD})^2}\\
   &+\frac{\left((\frac{\varepsilon}{T^4})_i^{LQCD}-(\frac{\varepsilon}{T^4})_i^{EVHRG}(R_h)\right)^2}{( (\Delta\frac{\varepsilon}{T^4})_i^{LQCD})^2}].
 \end{split}
\end{align}
In the last expression, $(\Delta\frac{P}{T^4})_i^{LQCD}$ and $(\Delta\frac{\varepsilon}{T^4})_i^{LQCD}$
are the errors of normalized pressure and energy density respectively at the $i$ th point
calculated in the LQCD and $N$ is the number of LQCD data points.

Figure \ref{fig:chi2} shows the variation of the $\chi^2$ with hardcore radius parameter of hadrons $R_h$.
Same radii for all the mesons and baryons have been considered in this work.
The upper limit of temperature for the continuum limit LQCD data ~\cite{Bazavov:2014pvz} is 
taken as $T = 200$ MeV to calculate $\chi^2$ of Fig. \ref{fig:chi2}.
It has been assumed that below $T = 200$ MeV HRG and EVHRG models are valid to describe
thermodynamic quantities because for this temperature equation states for both
ideal HRG and EVHRG are well below the Stefan-Boltzmann limit for the deconfined QGP phase. 
Horizontal lines in the Fig. \ref{fig:chi2} indicate the  $\chi^2_{min}$ and the $\chi^2_{min} + 1$ values.
The $\chi^2_{min}$ corresponds to the best fit of the LQCD data and the $\chi^2_{min} + 1$
indicate the errors~\cite{Andronic:2005yp, Andronic:2016nof} on the parameter $R_h$.
One can see from the Fig. \ref{fig:chi2}, the best fit in terms of $\chi^2$ is 
achieved at $R_h = 0.15^{+0.04}_{-0.08}$ fm with $\chi^2_{min} = 0.42$. 
This estimate is slightly smaller compared to the previous observations 
\cite{Andronic:2012ut, Bhattacharyya:2013oya}
where it was shown that most of the thermodynamic quantities can be described taking hadronic
radii between $0.2 - 0.3$ fm.
The value of the hardcore radius was estimated as $R_h = 0.3$ fm in the Ref ~\cite{BraunMunzinger:1999qy}.
Whereas in Ref. ~\cite{Yen:1997rv} $R_h = 0.3 -0.8$ fm was used to fit experimental hadronic yields.
Although this estimation of
$R_h$ in this present work depends on the range of temperature used for the fit it will give us some idea about
the $R_h$ of EVHRG model where only one extra parameter ($R_h$) is introduced compared to the ideal HRG model.

\begin{figure}[]
\centering
 \subfigure{\includegraphics[width=0.45\textwidth]{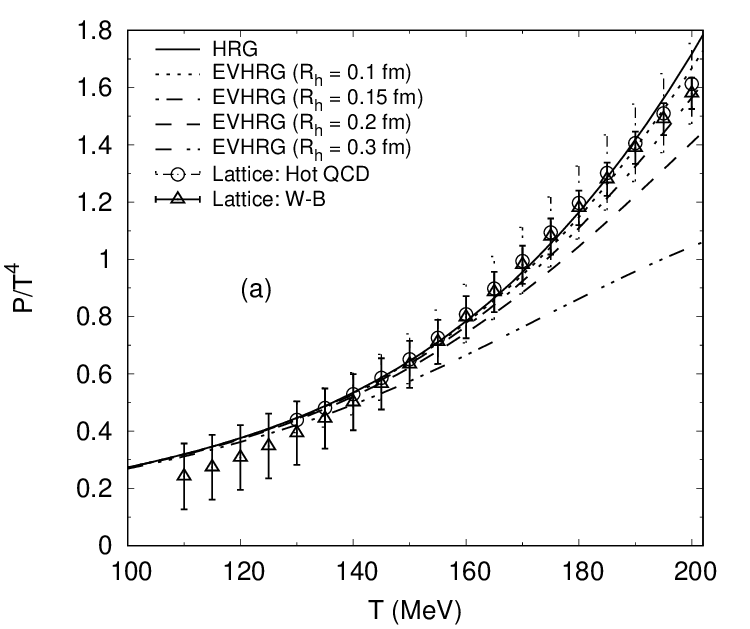}\label{P_T}}
\subfigure{ \includegraphics[width=0.45\textwidth]{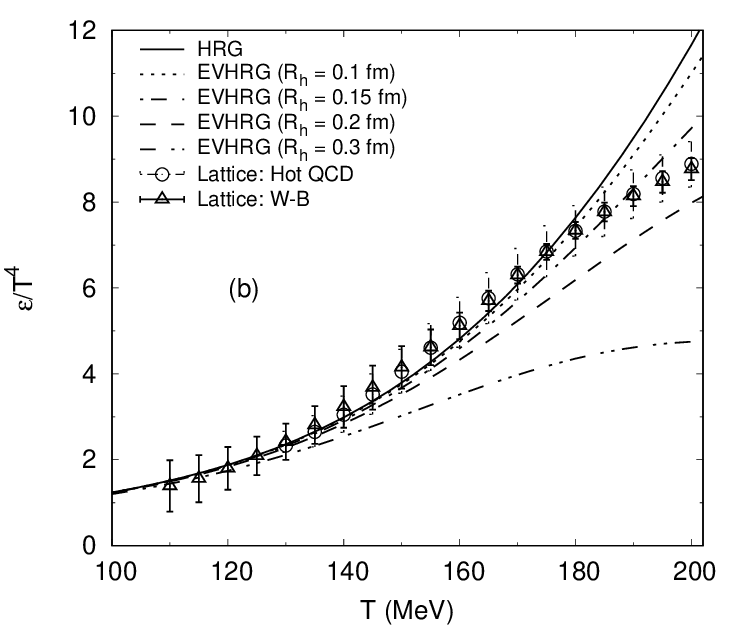}\label{E_T}}
\vspace{0.5cm}
\caption{Variation of normalized pressure (a) and energy density (b) with the temperature
at $\mu=0$.
Continuum limit lattice QCD data are taken from Hot QCD Collaboration ~\cite{Bazavov:2014pvz} and 
Wuppertal-Budapest Collaboration ~\cite{Borsanyi:2013bia}.}
 \label{fig:P_E_vs_T}
\end{figure}

Figure \ref{fig:P_E_vs_T} shows the temperature dependence of
normalized pressure and normalized energy density respectively at $\mu=0$. 
Results of ideal HRG model have been represented by solid lines.
Other lines correspond to the results of the interacting HRG model or the EVHRG model
with different radii of hadrons ($R_h$). 
It can be seen that there is almost no effect of interaction
till $T \simeq 120$ MeV both in pressure and energy density.
The reason for this is that the
effective degree of freedom of the system does not increase much up to this
temperature and therefore correction due to excluded volume is small.
Beyond $T = 120$ MeV a substantial change in these quantities has been observed.
It can be seen from this
figure that at large $T$ normalized pressure as well as normalized energy density are suppressed 
compared to the ideal HRG if we take non-zero hardcore radii of the hadrons.
This is expected since hadrons start interacting at large temperature where the hadronic
population is large. 
Further, suppression increases with the increase of radii of the hadrons.
The continuum limit
LQCD data of Hot QCD Collaboration ~\cite{Bazavov:2014pvz} and Wuppertal-Budapest
Collaboration ~\cite{Borsanyi:2013bia} have also been plotted in this figure.
Figure \ref{P_T} illustrates the excellent agreement between the ideal HRG and the lattice QCD calculations
of normalized pressure.
On the other hand, one can see from Fig. \ref{E_T} that normalized energy density
calculated in ideal HRG model is close to the LQCD data up to the crossover 
temperature $T_c$ ($154 \pm 9$ MeV) ~\cite{Bazavov:2014pvz} of the LQCD calculation.
However, for both pressure and energy density high temperature region
agrees well with EVHRG model for $R_h = 0.15$ fm which indicates that interaction is very
important to include in the HRG model especially at the high temperature region. 
At the low temperature region interaction is also there. However its effect is small.
It should be noted that $R_h = 0.15$ fm gives
the best fit of the LQCD data upto $T = 200$ MeV which is already shown
in the Fig. \ref{fig:chi2}. 
The increase of the hardcore radii of all the hadrons
further reduces the ability to reproduce both the LQCD results of 
normalized pressure and normalized energy density as can be
seen from the Fig. \ref{fig:P_E_vs_T}.
It is found in Ref. \cite{Satarov:2016peb} that the equation of state can also be described well
with baryonic radius $0.3 - 0.6$ fm assuming mesons to be point like. 
In Ref. \cite{Albright:2014gva} description of of equation of state 
even higher temperature region ($T > 200$ MeV) is also improved using a hybrid model of EVHRG and
the perturbative QCD. In this work a switching function is used to connect hadronic and the 
partonic phase.
The mass dependent hadronic radius is also considered in EVHRG
model to study the hadronic yields at LHC energy and good agreement between model and 
experimental data is found \cite{Alba:2016hwx}.

\begin{figure}[]
\centering
 \subfigure{\includegraphics[width=0.45\textwidth]{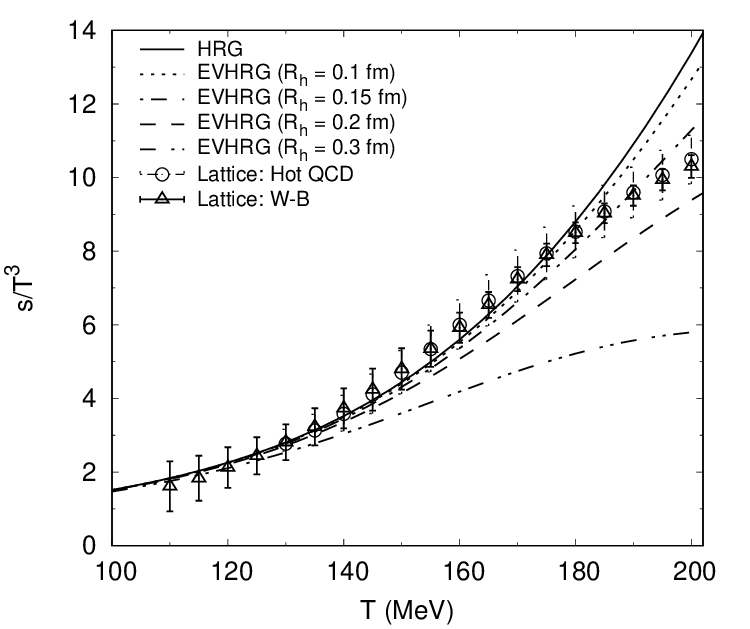}}
 \vspace{0.5cm}
\caption{The variation of normalized entropy density with the temperature at $\mu=0$.
Continuum limit lattice QCD data are taken from Hot QCD Collaboration ~\cite{Bazavov:2014pvz} 
and Wuppertal-Budapest Collaboration ~\cite{Borsanyi:2013bia}.}
 \label{fig:s_vs_T}
\end{figure}
%

Figure \ref{fig:s_vs_T} shows the variation of normalized entropy density with the temperature at $\mu=0$.
Similar to Fig. \ref{fig:P_E_vs_T}, results of $s/T^3$ calculated in the EVHRG model with 
$R_h = 0.15$ fm are very
close to the LQCD data. For larger radii $s/T^3$ of EVHRG model under estimate the LQCD results.

%

\subsection{\label{sec:cv_mu0}Specific heat at $\mu =0$}
The specific heat at constant volume $C_V$ is given by ~\cite{Bazavov:2014pvz}
\begin{align}
 \begin{split}
&C_V=\left(\frac{\partial \varepsilon}{\partial T}\right)_V.
 \end{split}
\end{align} 
The $C_V$ is a sensitive indicator of
the transition from hadronic matter to the QGP.
The $C_V$ increases rapidly or even diverges
near the transition temperature for a conventional second order phase transition.
Although results of pressure, energy density and entropy density in EVHRG model 
are known already from published literature,
result of $C_V$ in EVHRG model is shown first time in this paper.

\begin{figure}[]
\centering
 \includegraphics[width=0.45\textwidth]{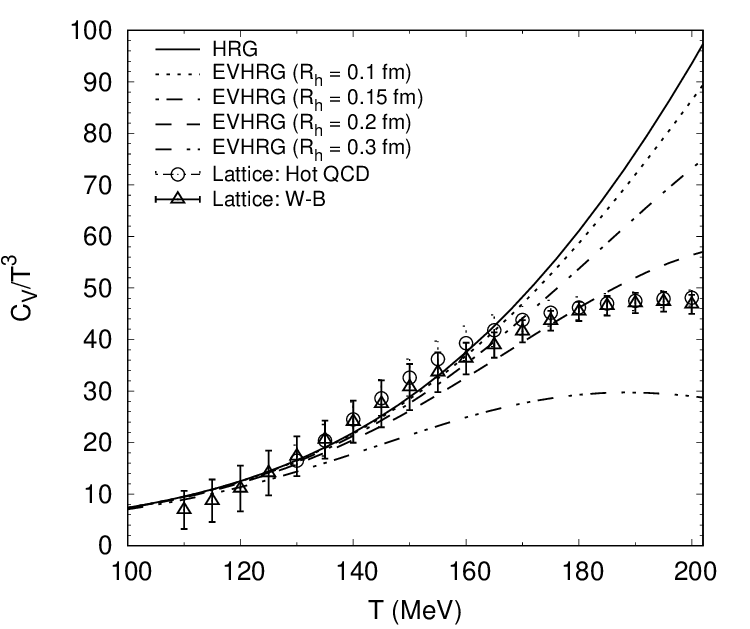}
 \vspace{0.5cm}
\caption{Variation of normalized specific heat at constant volume with temperature at $\mu=0$.
Lattice QCD data for continuum extrapolation are taken from Bazavov {\it et al.} ~\cite{Bazavov:2014pvz} and 
Borsanyi {\it et al.} ~\cite{Borsanyi:2013bia}.}
 \label{fig:Cv_vs_T}
\end{figure}

Figure \ref{fig:Cv_vs_T} shows the temperature dependence of normalized specific heat at $\mu=0$.
Similar like normalized pressure and energy density, normalized specific heat calculated in ideal HRG 
model is very close to the continuum limit LQCD data ~\cite{Bazavov:2014pvz, Borsanyi:2013bia} 
up to the temperature $T \simeq T_c$.
Results of normalized specific heat calculated in EVHRG model has also been shown in this figure.
Normalized specific heat in EVHRG model is suppressed compared to ideal HRG model and the 
suppression increases with increasing temperature as well as with increasing radii of hadrons
because of the repulsive interaction between hadrons.
The $C_V/T^3$ calculated in the EVHRG model with $R_h = 0.15$ fm is very
close to the LQCD data upto $T = 170$ MeV.
For $R_h = 0.3$ fm $C_V/T^3$ of EVHRG model matches only qualitatively with LQCD data
but not quantitatively.

So from Fig. \ref{fig:P_E_vs_T} - \ref{fig:Cv_vs_T}, it has been observed
that results of EVHRG model are close to ideal HRG model only at low temperature region.
The effect of interaction is not avoidable at high temperature region.
However, to avoid complicacy
the modeling of the HRG as the ideal gas was considered in most of the works.

\subsection{\label{sec:p_cv_cfo}$P/T^4$ and $C_V/T^3$ at chemical freeze-out}

\begin{figure}[]
\centering
 \includegraphics[width=0.45\textwidth]{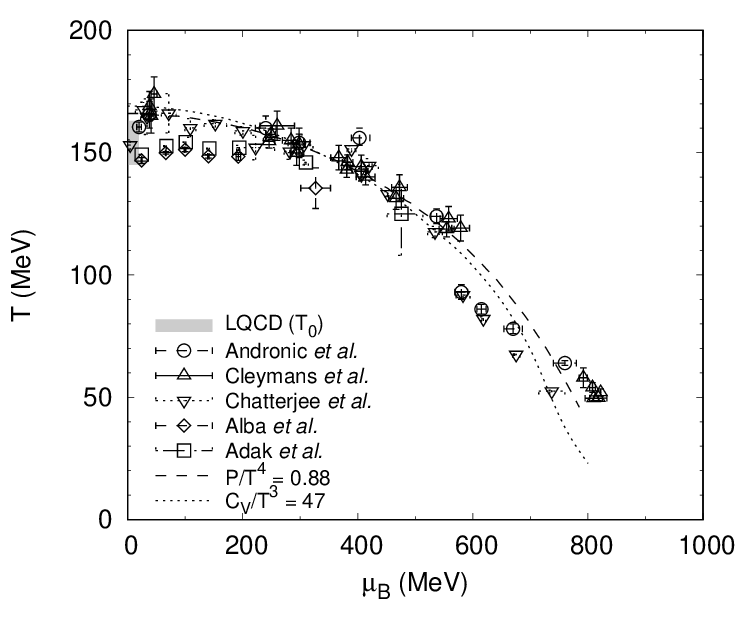}
 \vspace{0.5cm}
\caption{Chemical freeze-out parameters ($T, \mu_B$) obtained 
by different 
groups ~\cite{Andronic:2005yp, Cleymans:2005xv, Chatterjee:2015fua,Alba:2014eba,Adak:2016jtk}
at various $\sqrt{s_{NN}}$ along with the
line of $P/T^4 = 0.88$ and $C_V/T^3 = 47$ calculated in the ideal HRG model. 
The blue box shows the uncertainty in LQCD transition temperature ($154 \pm9$ MeV)
at $\mu = 0$ \cite{Bazavov:2011nk}.}
 \label{fig:freezeout_line_p_cv}
\end{figure}

\begin{figure}[]
\centering
 \includegraphics[width=0.45\textwidth]{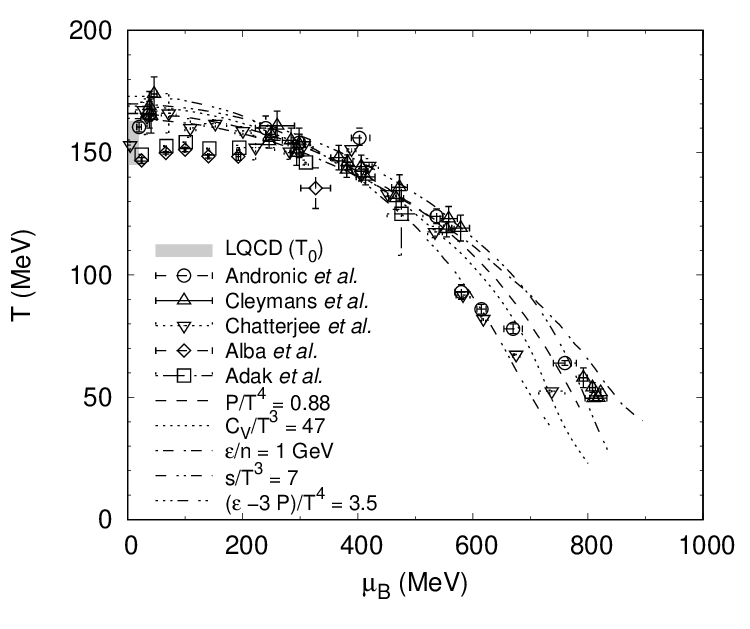}
 \vspace{0.5cm}
\caption{Same as Fig. \ref{fig:freezeout_line_p_cv} but with additional
comparisons with other constant chemical freeze-out lines 
calculated in the ideal HRG model.}
 \label{fig:freezeout_line_all}
\end{figure}

\begin{figure}[]
\centering
\includegraphics[width=0.45\textwidth]{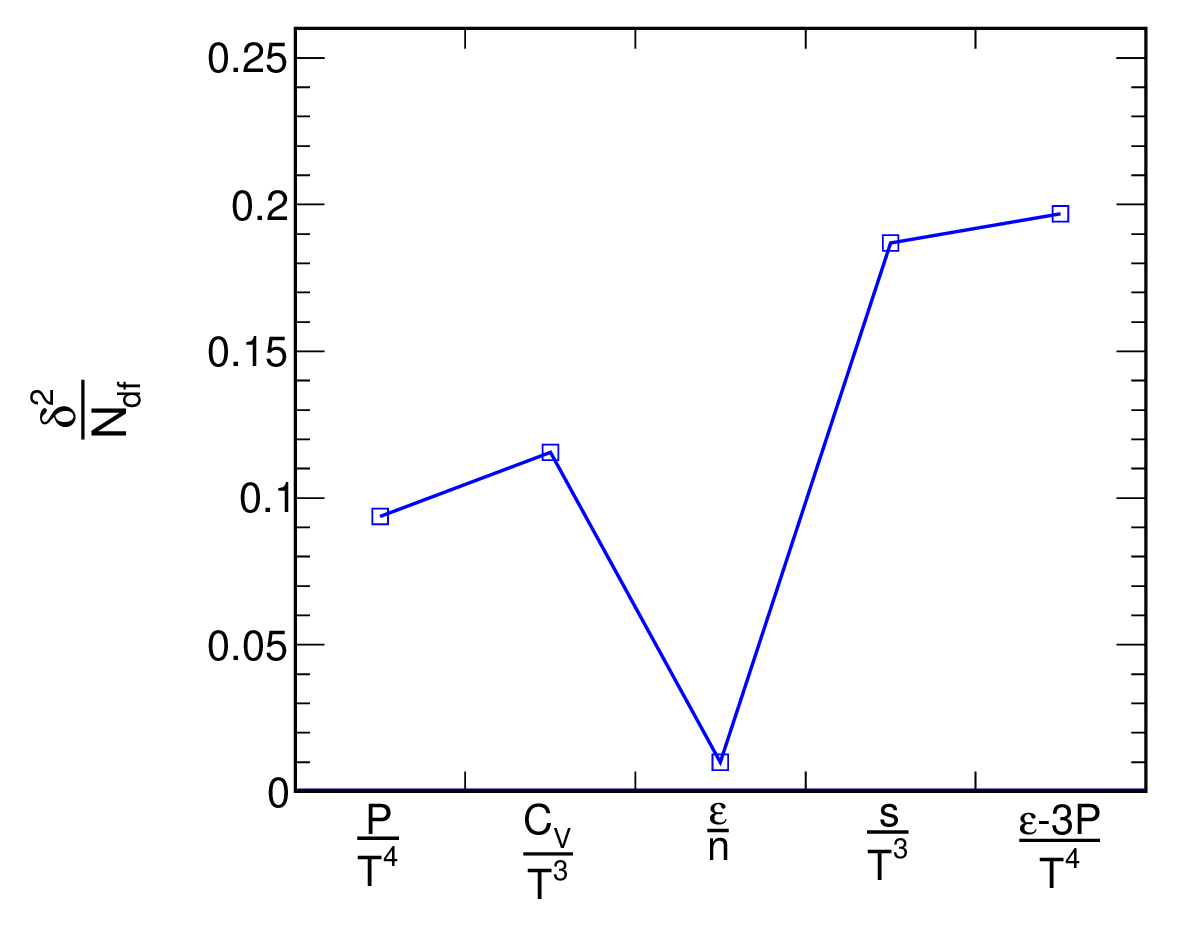}
\vspace{0.5cm}
\caption{The deviation, $\delta^2$ per degrees of freedom for different observables related to
the universal chemical freeze out lines shown in Fig. \ref{fig:freezeout_line_all}.}
\label{fig:deviation}
\end{figure}

The thermal fireball created due to heavy ion collision expands and cools.
At the initial stage, a large number of particles are produced due to deposition of a 
huge amount of energy in the core of the collision.
At this stage, particles collide mostly inelastically.
After some time inelastic collisions among the particles stop and hence hadronic yields or ratios get fixed.
This stage is called chemical freeze-out. It is already mentioned in the Sec. \ref{sec:Intro} that, from the experimental information 
about hadronic ratios or hadronic yields chemical freeze-out parameters can be calculated phenomenologically
~\cite{Cleymans:2005xv,Xu:2001zj,Becattini:2005xt,Andronic:2005yp,Andronic:2009jd,
Karsch:2010ck,Chatterjee:2015fua}. Although, all those calculations ignored any dynamics of the system.
Chemical freeze-out parameters can also be calculated phenomenologically
from experimental data of fluctuations ~\cite{Alba:2014eba,Adak:2016jtk}.
There are several universal chemical freeze-out descriptions in the existing literature
which give quite satisfactory descriptions of the hadronic multiplicities
measured in heavy-ion collisions.
These universal chemical freeze-out properties include
$\varepsilon/n=1 $ GeV~~\cite{Cleymans:1998fq,Cleymans:2005xv},
$s/T^3=7$~~\cite{Tawfik:2004vv},
$n_B + n_{\bar{B}} = 0.12$ fm$^{-3}$~~\cite{BraunMunzinger:2001mh},
$(\varepsilon-3P)/T^4 = 7/2$ ~\cite{Tawfik:2013eua} etc., where
$n_B$ is the baryon density and $n_{\bar{B}}$
is the anti-baryon density.
In this paper, two more universal descriptions of chemical freeze-out
have been proposed, namely $P/T^4 \simeq 0.88$ and $C_V/T^3 \simeq 47$.

Figure \ref{fig:freezeout_line_p_cv} shows chemical freeze-out parameters
in ($T, \mu_B$) plane calculated in HRG model by different 
groups ~\cite{Andronic:2005yp, Cleymans:2005xv, Chatterjee:2015fua,Alba:2014eba,Adak:2016jtk}
along with $P/T^4= 0.88$ and $C_V/T^3 \simeq 47$ calculated in the ideal HRG model.
Different symbols with error bars represent phenomenologically
calculated chemical freeze-out
parameters from experimental data at different $\sqrt{s_{NN}}$
ranging from a couple of GeV at SIS up to several TeV at LHC.
In the Fig. \ref{fig:freezeout_line_p_cv} dashed line shows $P/T^4= 0.88$
whereas dot line shows $C_V/T^3= 47$.
It can be seen that this constant normalized pressure can reproduce
chemical freeze-out parameters at various $\sqrt{s_{NN}}$ in ($T, \mu_B$) plane. 
Similarly, $C_V/T^3= 47$ calculated in the ideal HRG model reproduces very well the
chemical freeze-out diagram. The blue box in this figure shows the uncertainty in LQCD 
transition temperature ($T_0= 154 \pm9$ MeV) \cite{Bazavov:2011nk} at $\mu = 0$.
The $P/T^4= 0.88$ and $C_V/T^3= 47$ lines touch the temperature axis at slightly
higher temperatures than $T_0$.

Figure \ref{fig:freezeout_line_all} is same as Fig. \ref{fig:freezeout_line_p_cv} 
but with additional
comparisons with other constant chemical freeze-out lines 
calculated in the ideal HRG model.
It can be seen that the present work is consistent with the previous 
works related to universal chemical freeze-out descriptions 
~\cite{Cleymans:1998fq,Cleymans:2005xv,Tawfik:2004vv,Tawfik:2013eua,Tawfik:2016jzk}.
It is worth mentioning that the conservation laws given in Eqs. \ref{eq:ns}- \ref{eq:nbq} have been incorporated
during calculations of different observables in ($T, \mu_B$) plane.

To show the goodness of theoretical curves,
a quantity is defined as:
\begin{equation}
 \delta^2 = \sum_i  \frac{(O_u - O_i)^2 }{O_i^2},
\end{equation}
where sum is over the chemical freeze-out points, $O_i$ and $O_u$ are 
respectively the values of the observable at the $i$ th point and 
its universal (fixed) value described previously. For an example,
if $O_i$ is $P/T^4(T_i,\mu_i)$ then $O_u = 0.88$. 
Figure \ref{fig:deviation} shows the deviation, $\delta^2/N_{df}$ for different observables
where $N_{df}$ is the number of degrees of freedom $i.e.,$ the 
number of point less the number of parameters in the model. It can be seen that
$\varepsilon/n = 1$ GeV gives the best description of the chemical freeze-out parameters.
Further, deviations for $P/T^4 =0.88$ and $C_V/T3 = 47$ are less compared to that of
$s/T^3 =7$ and $(\varepsilon -3P)/T^4 = 3.5$.

\begin{figure}[]
\centering
 \subfigure{\includegraphics[width=0.45\textwidth]{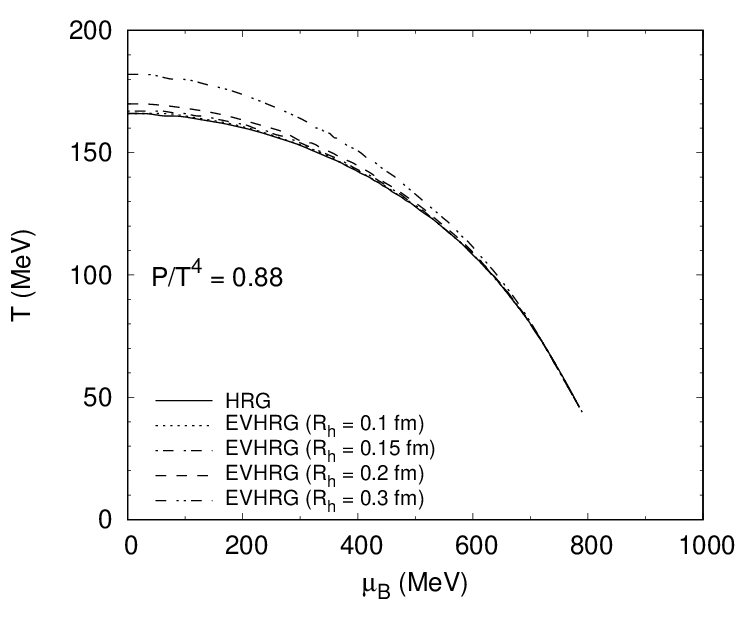}\label{P_fo_int}}
\subfigure{ \includegraphics[width=0.45\textwidth]{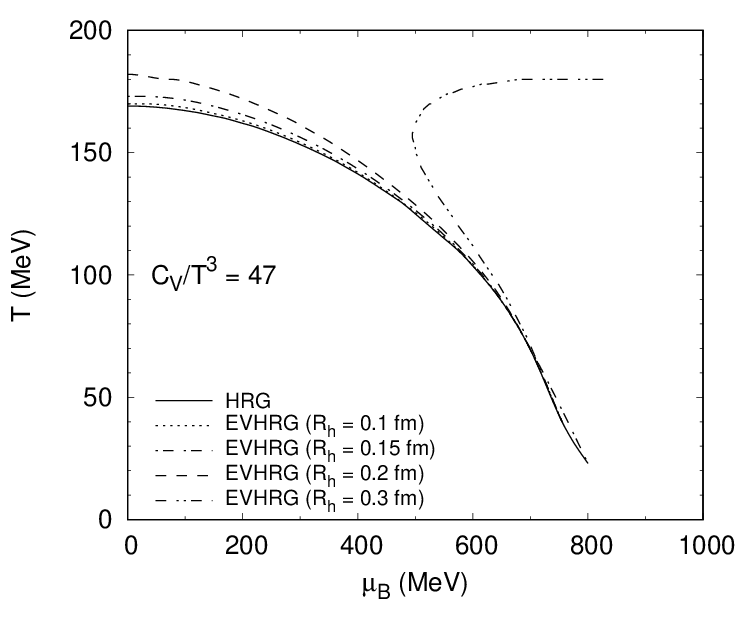}\label{cv_fo_int}}
 \subfigure{ \includegraphics[width=0.45\textwidth]{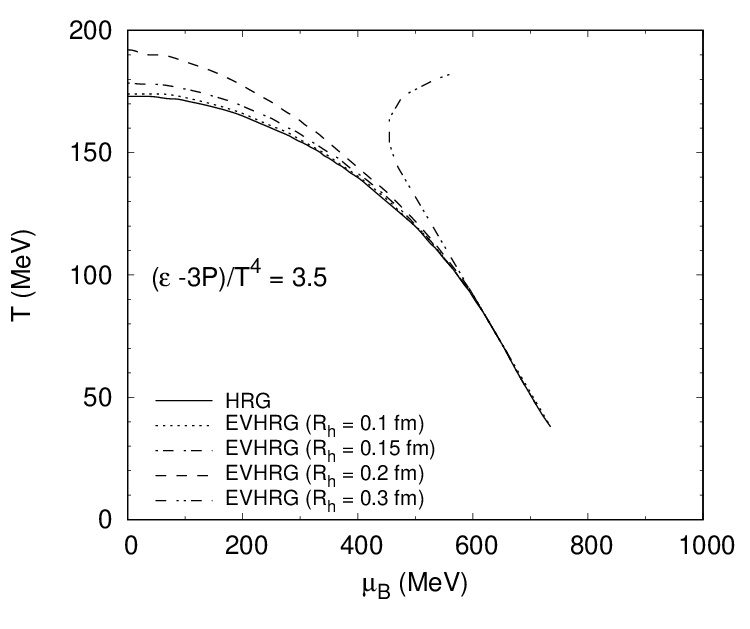}\label{tr_an_fo_int}}
\subfigure{ \includegraphics[width=0.45\textwidth]{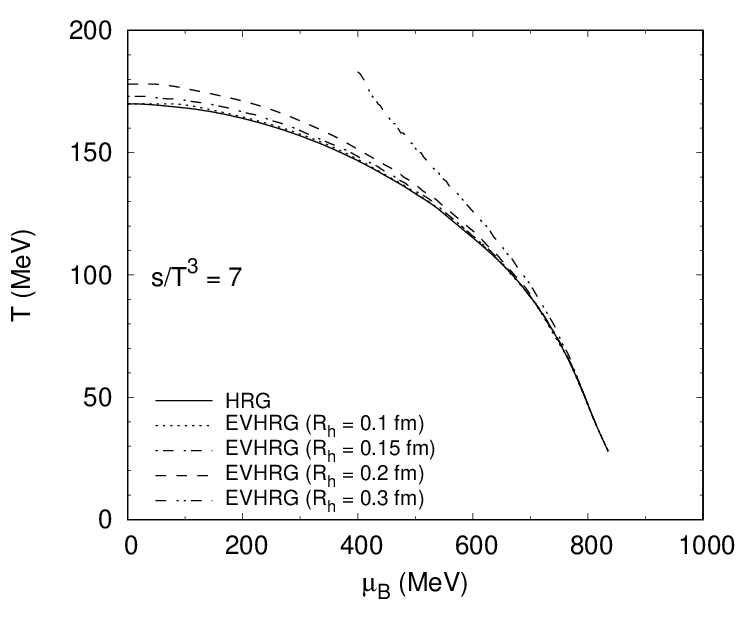}\label{s_fo_int}}
\subfigure{ \includegraphics[width=0.45\textwidth]{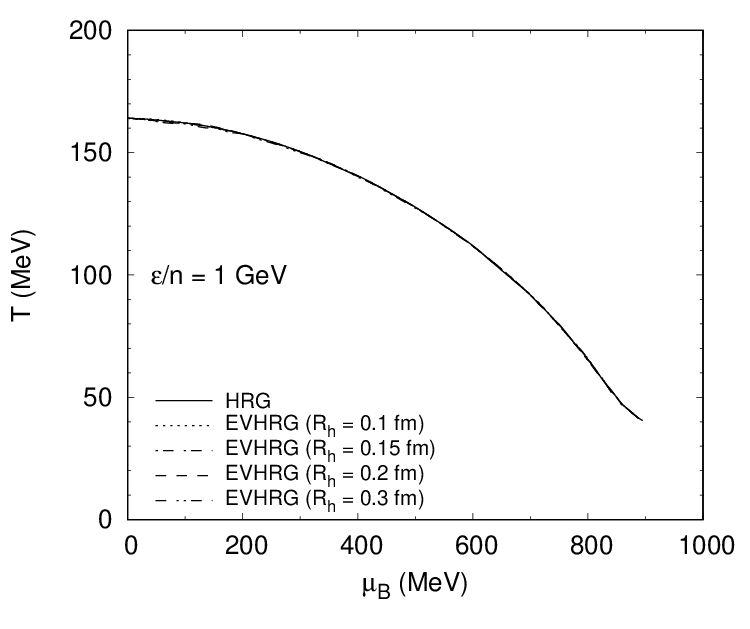}\label{e_n_fo_int}}
\vspace{0.5cm}
\caption{Variation of universal chemical freeze-out lines with hadronic radii $R_h$.}
 \label{fig:fo_int}
\end{figure}

Let us now discuss the physical interpretation of the proposed
universal chemical freeze-out lines. It is really remarkable to see that some basis
thermodynamic quantities do not change along the chemical freeze-out line over a
large energy range. This indicates the equilibration of the thermal matter 
even at lower collision energy. 
In the Boltzmann approximation, $P/T^4$ and $C_V/T^3$ can be written as
\begin{equation}
 \frac{P}{T^4} = \sum_i \frac{g_i z_i m_i^2}{2\pi^2T^2}K_2(\frac{m_i}{T}),
\end{equation}
\begin{equation}\label{cv_BS}
 \frac{C_V}{T^3} = \sum_i \frac{g_i z_i}{2\pi^2}\left(12\frac{m_i^2}{T^2} K_2(\frac{m_i}{T})
 +\frac{m_i^4}{T^4} K_2(\frac{m_i}{T})+3 \frac{m_i^3}{T^3} K_1(\frac{m_i}{T})\right).
\end{equation}
where $z_i = \exp(\mu_i/T)$ is the fugacity for the $i$ th hadron.
At very high energy collision (say LHC energy) chemical freeze-out $T$ does not vary much with the change
of $\sqrt{s_{NN}}$ and $z_i\sim1$ since $\mu_i$ becomes almost zero. Hence 
$P/T^4$ and $C_V/T^3$ remain fixed. 
At low energy the $P/T^4$ is effected by two factors; $z_i/T^2$ and the $K_2(1/T)$ for a fixed $m_i$.
With the decrease of collision energy chemical freeze-out $T$ decrease whereas $\mu_i$ increases.
Therefore $z_i/T^2$ increases and the other function $K_2(1/T)$ decreases.
As a result, $P/T^4$ of the hadron gas remains almost constant.
Similarly, with the decrease of collision energy each term of $C_V/T^3$ in Eq. \ref{cv_BS} remains
almost constant. 
In the hadronic matter pressure and energy density are 
related by the equation $P = c_s^2 \varepsilon$, where $c_s^2$ is the 
speed of sound of the medium.
Now the specific heat is related to the $c_s^2$ by the following equation \cite{Bazavov:2014pvz}
\begin{equation}
c_s^2 =\frac{\partial P}{\partial {\varepsilon}}=\frac{\frac{\partial P}{\partial T}}{\frac{\partial \varepsilon}{\partial T}} =\frac{s}{C_V} .
\end{equation}
For a massless, noninteracting gas, $c_s^2 = 1/3$. 
For a hadronic medium $c_s^2$ is expected to be less than that.
In Ref. \cite{Mohanty:2003va} it was shown that the hadronic spectra can be described well in Landau
hydrodynamical model with $c_s^2 = 0.2$.
In this present work we note that
along the chemical freeze-out line
$c_s^2$ is constant and has the value $\simeq 0.15$ ($\simeq 7/47$) 
since both $s/T^3$ and $C_V/T^3$ are constant along this line.
This tells that medium properties in terms of $c_s^2$ are also similar along the 
chemical freeze-out line $i.e.,$ in all collision energies.

\section{\label{sec:sensitivity}Sensitivity of chemical freeze-out lines on repulsive interaction}
Repulsive interaction affects the equation of states of the thermal system
at high temperature which is already discussed in this paper.
Therefore, it is necessary to 
study the sensitivity of universal chemical freeze-out lines in presence of
repulsive interaction.

Figure \ref{fig:fo_int} shows different universal chemical freeze-out conditions in EVHRG model for 
different radii of hadrons. There is no effect of repulsive interaction is there at chemical freeze-out
$\varepsilon/n=1 $ GeV condition. This is because suppression factor,
($1+\sum_k V_{ev,k}n_k^{id}(T,\hat{\mu_k})$)
is there at the denominator for both energy density and number density as can be seen from 
the Eqs.. \ref{eq:e_ev} -\ref{eq:n_ev}. For all other universal chemical freeze-out conditions
effect of repulsive interaction is clearly visible. 
However, for $R_h = 0.15$ fm at which $\chi^2$ is minimum
(Fig. \ref{fig:chi2}), universal chemical freeze-out lines deviate from the 
corresponding lines of ideal HRG 
only upto $3 \%$ but qualitative behaviors remain same. Although at very large radii 
(say $R_h = 0.3$ fm, for an example)
shapes of the universal chemical freeze-out lines for $C_V/T^3 = 47$, $(\varepsilon - 3 P)/T^4 = 3.5$
and $s/T^3 = 7$ become different compared to ideal HRG model.

\section{\label{sec:conclusions}Conclusions}

I conclude that at high temperature the ideal HRG model is not good enough to
describe LQCD data
of pressure, energy density, entropy density and specific heat calculated at $\mu=0$.
The EVHRG model with hadronic radii of $0.15$ fm gives best description of pressure, energy density 
where the minimum 
of $\chi^2$ has been observed. The same hadronic radii can describe LQCD data of $s/T^3$ upto $T = 200$ MeV
and of $C_V/T^3$ upto $T = 170$ MeV. 
All these results indicate the importance of repulsive interaction present in the EVHRG model.
The chemical freeze-out parameters deduced 
at various $\sqrt{s_{NN}}$ are well reproducible in the ideal HRG model
using the conditions of constant normalized pressure and constant normalized specific heat respectively.
It has been observed that both $P/T^4= 0.88$ and $C_V/T^3= 47$ calculated in ideal HRG model
reproduce very well the chemical freeze-out diagram which indicate that the basic thermodynamic
properties of the system created in heavy ion collision are almost similar in all collision energies.
Since $s/T^3$ is also constant along the chemical freeze-out curve,
$C_V/T^3= 47$ corresponds to $c_s^2 = 0.15$ of the medium.
Further, since repulsive interaction
is there in the hadronic medium values of $P/T^4$ and $C_V/T^3$ at chemical freeze-out
become model dependent. However for our best fit value of hadronic radii ($R_h = 0.15$ fm)
qualitative behaviors of $P/T^4$ and $C_V/T^3$ remain similar to ideal HRG model.
Hence $P/T^4$ and $C_V/T^3$ remain constant at chemical freeze-out even in EVHRG model.
In this present work pressure, energy density of the matter are used to extract the hadronic
radii. Susceptibilities of conserved charges can also be used for the same purpose.
This is beyond the scope of the present analysis and will be treated elsewhere.
In some very recent papers, van der Waals attractive interaction is also considered in the 
hadronic model 
~\cite{Vovchenko:2015xja, Vovchenko:2015vxa, Vovchenko:2015pya, Redlich:2016dpb, Vovchenko:2016rkn}. 
I plan to study the sensitivity of universal chemical freeze-out lines 
on the attractive interaction in my future work.

\section*{Acknowledgments}
I would like to thank Prof. S. K. Ghosh for useful discussions.

\end{document}